\documentclass[10pt,aps,twocolumn,nobibnotes,pra,showpacs]{revtex4-1}
  \usepackage{color}
  \usepackage{newlfont}
  \usepackage{graphicx}
  \usepackage{amssymb}
  \usepackage{amsmath}
  \usepackage[latin1]{inputenc}
  \usepackage{bbm}
  
\newcommand{\ket}[1]{\mbox{$ | #1 \rangle $}}
\newcommand{\bra}[1]{\mbox{$ \langle #1 | $}}
\newcommand{\be}{\begin{eqnarray}}
\newcommand{\ee}{\end{eqnarray}}

\begin{document}

\title{Bayes' rule, generalized discord, and nonextensive thermodynamics}
\author{A. C. S. Costa}
\author{R. M. Angelo}
\affiliation{Departamento de Física, Universidade Federal do Paraná, Caixa Postal 19044, 81531-980, Curitiba, PR, Brazil}

\date{\today}

\begin{abstract}
Generalized measures of quantum correlations are derived by taking Bayes' rule as the only fundamental principle. The resulting quantifiers satisfy several desirable conditions for a measure of quantum correlations and are shown to admit operational interpretation in terms of the difference in efficiency of quantum and classical demons in allowing for the extraction of generalized work from a heat bath. The link with discord is established by adopting the $q$ entropy as entropic principle. This allows us to reproduce, within a one-parameter formalism, both the entropic and the geometric measures of discord and physically distinguish them within the context of the nonextensive thermodynamics. Besides offering a unified view of several measures of correlations in terms of the Bayesian principle and its connection with thermodynamics, our approach unveils a bridge to the nonextensive statistical mechanics.
\end{abstract}

%03.65.-w	Quantum mechanics
%03.67.-a	Quantum information
%03.65.Ta	Foundations of quantum mechanics; measurement theory
%03.67.Mn	Entanglement measures, witnesses, and other characterizations
%\keywords{quantum integrability, dynamical invariants}

\maketitle

%============================================================
\section{Introduction}

Correlations occupy a prominent place in our present models of nature. Acquisition of empiric {\em information} from the world is typically mediated by the fingerprints left by these resources in physical pointers. The information stored in the apparatus, a physical memory which necessarily gets an increase of {\em entropy}, can eventually be used to extract {\em work} from a heat bath. The whole process is such that we can learn about and benefit from the laws of nature without violating any of them~\cite{landauer,vedralD}.

Entanglement figures in this context as a class of correlations which cannot be prepared by local operations and classical communication (LOCC)~\cite{h409}. It permeates many physical arenas, from foundational phenomena, such as nonlocality~\cite{popescu94} and decoherence~\cite{zurek03}, to the interplay with the quantum information science~\cite{wolf10,chuang}. About a decade ago, however, a novel class of quantum correlations was identified which manifests even in separable states. Since then, the so-called {\em quantum discord}~\cite{zurek01,vedral01}, along with some of its variants~\cite{vedral10,fu10,paris10,sarandy11}, has revealed its importance to a variety of scenarios, including approaches in quantum information~\cite{rajagopal08,williamson10,winter11,rigolin10,walther12,lam12}, quantum computation~\cite{caves08,white08,datta09,fan10}, broadcasting of quantum states~\cite{h208,piani09}, decoherent dynamics~\cite{boas09,caldeira10,serra10,maniscalco10,guo10}, critical systems~\cite{dille08,Wrigolin10}, and biology~\cite{uskov10} (see Refs.~\cite{celeri11,vedral12} for recent reviews on the subject). One of the versions of this information theoretic measure, here called {\em entropic discord} and stated as
\be
D_E(\rho)\!=\!\min\limits_{\{\Pi_y\}}\!\Big(S_1(\Pi_Y[\rho_Y])+\sum_yp_yS_1(\rho_{X|y}) \Big)\!-\!S_1(\rho),\,\,\,\,\,
\label{DE}
\ee
has been linked with the difference between the efficiency of quantum and classical demons in extracting thermodynamic work from a heat bath~\cite{celeri11,zurekD}. Above, $\rho_{X|y}=\text{Tr}_Y(\Pi_y\rho)/p_y$ is the conditional density matrix, $p_y=\text{Tr}(\Pi_y\rho)$ is the probability of the outcome $y$, $\Pi_Y[\rho]=\sum_y\Pi_y\,\rho\,\Pi_y$, $\Pi_y=\ket{y}\bra{y}$ is a von Neumann projector associated with a discrete observable $Y$, and $S_1$ is the von Neumann entropy. Alternative measures have been constructed via geometric principles. They have shown to be of great potential for both theoretical~\cite{celeri11,vedral12} and experimental investigations~\cite{laflamme12,adesso12}. In its seminal version, the {\em geometric discord}~\cite{vedral10} reads
\be
D_G(\rho):=\min\limits_{\Pi_Y}||\,\rho-\Pi_Y[\rho]\,||^2,
\label{DG}
\ee
where $||\rho||^2:=\text{Tr}(\rho^{\dag}\rho)$ is the square norm in the Hilbert-Schmidt space. To date, there is no conceptual framework elucidating the physical difference, if any, between the entropic and the geometric discord.

The quest for a fine understanding of these and other measures of quantum correlations associated with measurement-induced disturbance~\cite{luo08,ciliberti10,adesso11jpa} and their eventual connections with thermodynamics is a formidable current problem in quantum physics~\cite{zurekD,terno10,vedralD}. In particular, it is legitimate to ask whether the available measures capture different quantum correlations or are just alternative mathematical expressions of the same resource. Whichever the case, it is insightful to look for unifying principles capable of revealing the actual substance of a set of measures. This paper aims to give some contributions in this direction. First, a generalized measure of quantum correlations is derived from a primitive notion of the Bayesian theory of probabilities. Second, an operational interpretation is given which relates our measure with the extra amount of information a quantum demon can provide from a state in comparison with a classical demon. Finally, by adopting a specific entropic principle, we obtain a unified view for the entropic and geometric measures of discord and establish an interpretation for them within the context of the nonextensive thermodynamics.

%---------------------------------------------------------------
\section{Bayesian correlations}

At the core of the classical theory of probabilities is the notion of conditional probability, $\wp_{X|y}:=\wp_{X,y}/\wp_y$. (Following Ref.~\cite{zurek01} we refer to this expression as {\em Bayes' rule}.) According to the Bayesian interpretation, $\wp_{X|y}$ refers to knowledge available about a random variable $X$ {\em after} a given outcome $y$ has been obtained in a measurement of a random variable $Y$. On the other hand, $\wp_{X,y}$ and $\wp_y$, respectively, denote the probability distribution of $X\cap (Y\!=\!y)$ and the probability of the outcome $y$, with no reference to measurements. These terms can be regarded as knowledge stored {\em before} the inference process. The {\em Bayesian updating} induced on a marginal probability distribution $\wp_{X}$ by the evidence $y$ about $Y$ can be expressed as
\be
\wp_{X}\stackrel{y}{\longrightarrow}\wp_{X|y}=\wp_{X,y}/\wp_y.\nonumber
\label{BC}
\ee
It is interesting to note that while the acquisition of data does imply updating of information in a single run of the experiment, as $\wp_X\neq \wp_{X|y}=(\wp_{X,y}/\wp_X\wp_y)\wp_X$ whenever $X$ and $Y$ are dependent events, there is no net effect on average, since $\sum_y\wp_y\wp_{X|y}=\sum_y\wp_{X,y}\equiv\wp_X$. 

Now, in the quantum context it is inescapable to link Bayesian updating with the quantum {\em collapse} (see Ref.~\cite{fuchs} for a related discussion). Under a projective measurement $\Pi_y$, the postulate of the reduction requires the reduced state to be updated as
\be
\rho_X \stackrel{y}{\longrightarrow} \rho^{collapsed}_{X}(y)=\rho_{X|y}=\text{Tr}_Y(\Pi_y\rho)/p_y.\nonumber
\label{BQ}
\ee
Here as well, there is effective updating only in single events, since in general one has that $\rho_X=\text{Tr}_Y\rho\neq\rho_{X|y}$ but $\sum_yp_y\rho_{X|y}=\rho_X$ (unread measurements). The analogy suggests that the quantum collapse can be interpreted as a subjective Bayesian updating of information rather than an objective physical process. There is, though, a subtle aspect that prevents the analogy to be complete: the {\em quantumness of the correlations}. To better illustrate the point, in what follows we devise a generic construction that avoids, at a first stage, the use of traditional quantities of the information theory, such as the mutual information, and thus highlights the violation of the Bayesian principle in quantum mechanics. 

We start by considering an arbitrary {\em continuous} function $f:\mathbb{R}\mapsto\mathbb{R}$. The application of $f$ on Bayes' rule $\wp_y\wp_{x|y}=\wp_{x,y}$ allows one to write a trivial equality,
\be
\sum_{x,y}f(\wp_y\wp_{x|y})=\sum_{x,y}f(\wp_{x,y}),
\label{DBC}
\ee
whereby we see that the right-hand side essentially refers to knowledge prior to measurements on $Y$.
[For concreteness the reader may imagine $f(\wp_{X,Y})$ as some entropic measure for the information associated with the distribution $\wp_{X,Y}$.] A quantum mechanical analog of this formula can be tried as follows. Noticing that $\Pi_Y[\rho]\,\Pi_y=p_y\rho_{X|y}\Pi_y$ and $\sum_y\Pi_y=\mathbbm{1}_Y$ we can immediately check that $f(\Pi_Y[\rho])=\sum_yf(p_y\rho_{X|y})\Pi_y$. Using this relation we get
\be
\text{Tr}_X\sum_{y}f\big(p_y\rho_{X|y}\big)=\text{Tr}f\big(\Pi_Y[\rho]\big).
\label{QBR}
\ee
Now, while in the right-hand side of Eq.~\eqref{DBC} there is no reference to measurements, in Eq.~\eqref{QBR} the influence of projective measurements explicitly manifests via $\Pi_Y[\rho]$. Nevertheless, this is not so for all states. In fact, if $\rho=\Pi_Y[\sigma]$, a {\em quantum-classical} state, then $\text{Tr}f(\Pi_Y[\rho])=\text{Tr}f(\rho)$ and the analogy is fully recovered between Eqs.~\eqref{DBC} and \eqref{QBR}. For general states, however, the analogy with the Bayesian updating fails. This motivates us to define a quantifier of the {\em least deviation from Bayes' rule} induced by local measurements,
\be
\Delta B(\rho) := \min\limits_{\Pi_Y}\text{Tr}\Big(f(\rho)-f(\Pi_Y[\rho])\Big)\geqslant 0.
\label{DB}
\ee
Non-negativity is readily demonstrated for any {\em differentiable convex} $f$ by means of the generalized Klein's inequality~\cite{carlen} along with the relation $\text{Tr}[\rho \,g(\Pi_Y[\rho])]=\text{Tr}[\Pi_Y[\rho]\,g(\Pi_Y[\rho])]$, which holds for any function $g$. As a byproduct of the proof we obtain the upper bound
\be
\Delta B^{ub}(\rho)=\min\limits_{\Pi_Y}\text{Tr}\Big\{\big(\rho-\Pi_Y[\rho]\big)\,f'(\rho)\Big\}.
\label{DBUB}
\ee
Additionally, if we require $f$ to be {\em strictly convex}, then the analogy between collapse and Bayesian updating will apply ($\Delta B(\rho)=0$) {\em iff} the state does not change under local measurements, i.e., $\Pi_Y[\rho]=\rho$. Now, sensitivity to local disturbance, non-negativity, and mathematical structure intimately related with the trace distance, are symptoms of quantifiers of quantum correlations. Furthermore, following Ref.~\cite{modi12}, one may show (see the Appendix) that $\Delta B$ satisfies other desirable conditions, such as $\Delta B(\rho_X\otimes\rho_Y)=~0$ (no quantum correlation for product states), $\Delta B(U\rho\,U^{\dag})=\Delta B(\rho)$ for $U=U_X\otimes U_Y$ (invariance under local unitary operations), and continuity under small perturbations on $\rho$. Thus, measure \eqref{DB} emerges as a possible precursor of discord and other measures of quantum correlations associated with measurement-induced disturbances. In what follows we show that this is indeed the case. First, however, we assess the physical meaning of $\Delta B$.

%--------------------------------------------------------
\section{Thermodynamic interpretation}

So far we have assumed that $f$ is a strictly convex function. Thus, in contraposition to entropy, usually taken as a {\em concave} measure of ignorance, it is natural to associate $f$ with {\em information} $(\mathcal{I})$. Following Ref.~\cite{h305} we propose that
\be
\text{Tr}f(\rho)=\mathcal{S}_{max}-\mathcal{S}(\rho)\equiv \mathcal{I}(\rho),
\label{I}
\ee 
where $\mathcal{S}(\rho)$ denotes an arbitrary entropic measure and $\mathcal{S}_{max}$ is a constant used to set $\mathcal{I}=0$ for maximally mixed states. Given that $\Delta B\!=\!\mathcal{I}(\rho)\!-\!\max_{\Pi_Y}\mathcal{I}(\Pi_Y[\rho])$ we can now elaborate on traditional demonic protocols~\cite{zurekD}. 

Charlie wants to bargain with Maxwell's demons in order to get information about a given state $\rho$. The available demonic beings, however, never reveal the outcome of a measurement, but only the observable measured. Charlie then examines two classes of spirits. He knows that a {\em classical demon} can only perform local operations on the system. After correlating the system $X$ with an eigenbasis $\{\ket{y}\}$ of his physical apparatus, the demon could read off the pointer $Y$, say at position $y$, and predict the state $\rho_{X|y}\Pi_y$. Without information about the outcome, Charlie would have to average over all possibilities, thus accessing only a partial amount  $\mathcal{I}(\sum_yp_y\rho_{X|y}\Pi_y)$ of information. Being lucky enough, Charlie might invoke a demon which always chooses the optimal basis, in which case he would benefit from $\max_{\Pi_Y}\mathcal{I}(\Pi_Y[\rho])$. On the other hand, a {\em quantum demon} can perform measurements in global bases corresponding to observables that commute with the state of the system $XY$. Having learned about the measured observable, Charlie could infer $\rho$ and thus accumulate an amount $\mathcal{I}(\rho)$ of information. Charlie then concludes that quantum demons are more effective than classical ones as the former can offer an amount $\Delta B$ of extra information. (In order not to violate thermodynamic laws, Charlie needs to erase the memory of the demon after completion of the service.)

The link with thermodynamics can be established by noting that the informational content of $\rho$ allows Charlie to draw from a heat bath of temperature $\mathcal{T}$ an amount of work $\mathcal{W}(\rho)=k\mathcal{T} \mathcal{I}(\rho)$~\cite{zurekD,h305}. Although the physical link between work and information is unquestionable~\cite{landauer,vedralD}, that specific functional relation relies on the existence of a thermodynamic structure for generalized quantities $\{k,\mathcal{T},\mathcal{S},\mathcal{W}\}$ preserving the form of the usual laws derived from the Boltzmann-Gibbs-von Neumann entropy $S_1$. 
Assuming that this is the case, we get
\be
\mathcal{W}(\rho)-\max\limits_{\Pi_Y}\mathcal{W}(\Pi_Y[\rho])=k\mathcal{T}\Delta B(\rho).
\label{WB}
\ee
This result points out that the extra work a quantum demon allows Charlie to extract from a heat bath of temperature $\mathcal{T}$, by use of a state $\rho$, is fundamentally determined by how much the unread local measurement $\Pi_Y$ violates Bayes' rule. Besides being immediately applicable to standard thermodynamics, Eq.~\eqref{WB} holds true in more general settings, as shown next.

%--------------------------------------------------------
\section{Relation to quantum discord}

The connection of our main results \eqref{DB} and \eqref{WB} with other measures of quantum correlations is established by specializing the entropic principle in Eq.~\eqref{I}. Within the framework of unified $(q,r)$ entropies~\cite{ye06,rastegin}, the class arising for $r=1$ is of particular interest here. It refers to the Tsallis $q$ entropy~\cite{tsallis88,raggio95},
\be
S_q(\rho):=\frac{1-\text{Tr} \rho^q}{q-1}\quad (q>0\in \mathbb{R}),
\label{Sq}
\ee
which is {\em nonnegative} and {\em strictly concave}. It reduces to the von Neumann entropy (logarithm given in natural base) as $q\to 1$ and recovers the linear entropy $S_2=1-\text{Tr}\rho^2$ as $q=2$. We are now in position to define a generalized discord, $D_q(\rho):=\left[\Delta B(\rho)\right]_{S_q}$, as a specialization of the Bayesian measure induced by the Tsallis entropy. From Eqs.~\eqref{DB}-\eqref{I} and \eqref{Sq} we obtain
\be
D_q(\rho)=\min\limits_{\Pi_Y}\Big(S_q(\Pi_Y[\rho])-S_q(\rho)\Big),
\label{Dq}
\ee
with upper bound
\be
D_q^{ub}(\rho)=\min\limits_{\Pi_Y}~\frac{q}{q-1}\text{Tr}\Big(\rho^q-\Pi_Y[\rho]\,\rho^{q-1}\Big).
\label{UB}
\ee
Extensions to other $(q,r)$ entropies can be carried on straightforwardly. The positivity of $D_q$ ($\forall\,q> 0$) implies that the $q$ entropy cannot decrease under local von Neumann measurements, i.e., $S_q(\Pi_Y[\rho])\!\geqslant\!S_q(\rho)$. This is an important advantage to other formulations~\cite{plastino12}. Concerning the thermodynamic meaning of the $q$ discord, it is remarkable that Refs.~\cite{plastino01,rajagopal03} allow us to directly write Eq.~\eqref{WB} as $\mathcal{W}_q(\rho)-\max_{\Pi_Y}\mathcal{W}_q(\Pi_Y[\rho])=k_q\mathcal{T}_qD_q(\rho)$, for $q\in (0,2)$ and proper definitions for the $q$ work $\mathcal{W}_q$ and generalized Lagrange multiplier $\beta_q=(k_q\mathcal{T}_q)^{-1}$.

The relation of the $q$ discord \eqref{Dq} with $D_E$ and $D_G$ can be verified as follows. By use of the decomposition that precedes Eq.~\eqref{QBR} it is straightforward to show that $\text{Tr}[(\Pi_Y[\rho])^q]=\sum_yp_y^q\text{Tr}_X[(\rho_{X|y})^q]$. In addition, one can take the relation $[\text{Tr}_Y(\Pi_y\rho)]^q=\text{Tr}_Y[(\Pi_y\rho\Pi_y)^q]$, $\forall\,q>0$, to show that $\sum_yp_y^q=\text{Tr}_Y[(\Pi_Y[\rho_Y])^q]$. Combining these results we can prove that
\be
S_q(\Pi_Y[\rho])=S_q(\Pi_Y[\rho_Y])+\sum_yp_y^qS_q(\rho_{X|y}),\nonumber
%\label{TEC}
\ee
which is the $q$ version of the {\em joint entropy theorem}~\cite{chuang}. Equation \eqref{Dq} is then rewritten as
\be
\!\!\!\!\!D_q(\rho)=\min\limits_{\Pi_Y}\!\Big(\!S_q(\Pi_Y[\rho_Y])+\sum_yp_y^qS_q(\rho_{X|y})\!\Big)\!-S_q(\rho),\quad
\label{Dqalt}
\ee
where $S_q(X|Y):=\sum_yp_y^qS_q(\rho_{X|y})$ is the Tsallis conditional entropy. It is now obvious by Eqs.~\eqref{DE} and \eqref{Dqalt} that $D_{q\to 1}(\rho)\!=\!D_E(\rho)$. The link with the geometric discord is less apparent but easily shown as well. The natural guess is to look at $q=2$, as in this case we have a direct link between entropy and geometry, i.e., $1-S_2(\rho)=||\rho||^2$. Using the identity given right before Eq.~\eqref{DBUB} we get
\be
||\rho-\Pi_Y[\rho]||^2=S_2(\Pi_Y[\rho])-S_2(\rho),\nonumber
\ee
which proves that $D_{q=2}(\rho)=D_G(\rho)$. Geometric discord turns out to be, therefore, a measure of the deviation from Bayes' rule modeled by the linear entropy. 

The above results provide a unified view of discord as Bayesian correlations induced by $q$-entropies. It is worth noticing that further measures can be conceived by taking Bayes' theorem, $\wp_{x,y}=\wp_y\wp_{x|y}=\wp_x\wp_{y|x}$. Following our previous procedure one may propose, for example, $\delta B:=\min_{\Pi}\text{Tr}[f(\rho)-f(\Pi[\rho])]$, with $\Pi[\rho]=\sum_{x,y}\Pi_{xy}\rho\Pi_{xy}$ and $\Pi_{xy}=\Pi_x\otimes\Pi_y$, as a measure of joint disturbance. It would be interesting to assess how $\delta B$ and other possible formulations compare to usual global quantifiers, such as {\em AMID}~\cite{adesso11jpa} and {\em quantum deficit}~\cite{h305}.

%--------------------------------------------------------------------------------------------
\section{Examples}

\begin{figure}[ht]
\vskip2mm
\includegraphics[scale=0.367]{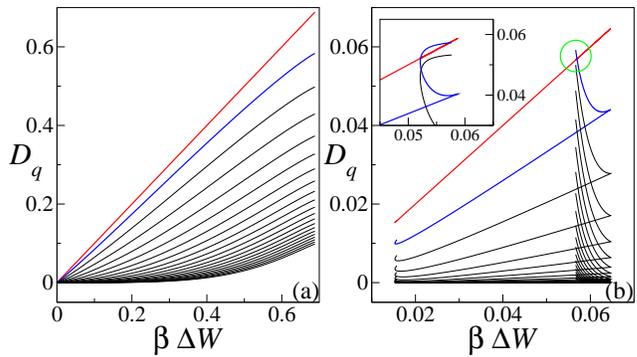}
\caption{(Color online) Parametric plot of the $q$-discord versus the excess of thermodynamic work $D_1=\beta\, \Delta W$ offered by quantum demons by use of (a) the Werner state $\vec{c}=-v(1,1,1)$ and (b) the $uv$-state $\vec{c}=(u,v,\frac{u-v}{2})$, with $u=1/3$, as a function of $v\in [0,u+\frac{2}{3}]$. From the top to the bottom, $q=1+n/2$ with $n=0$ (red), 1 (blue), $\dots$, 20 (lower black line). A similar plot is shown in the inset for $u=0.32$ and $n=0,1,2$. The $uv$-state is entangled if $v>(2-u)/3$.}
\label{fig1}
\end{figure}

For any pure state $\rho=\ket{\psi}\bra{\psi}$ one can take $\ket{\psi_{X|y}}\equiv\langle y|\psi\rangle/\sqrt{p_y}$ to show that $\text{Tr}_X\bra{y}\rho\ket{y}^q=p_y^q$. It follows that $\text{Tr}[(\Pi_Y[\rho])^q]\!=\!\text{Tr}_Y[(\Pi_Y[\rho_Y])^q]\!\leqslant\!\text{Tr}_Y\rho_Y^q$, the inequality deriving from $S_q(\Pi_Y[\rho_Y])\geqslant S_q(\rho_Y)$~\cite{rastegin}. Noting that $\text{Tr}\rho^q=1$ we then get
\be
D_q(\ket{\psi})=S_q(\rho_Y)\leqslant \frac{q}{q-1}S_2(\rho_Y)\quad(\forall q>0).\nonumber
%\label{Dqpure}
\ee
The upper bound was computed via Eq.~\eqref{UB}.
Supported by Ref.~\cite{sanders11} the previous expression extends to all $q>0$ the result according to which discord reduces to entanglement for pure states. In particular, for a maximally entangled state, $\ket{\Psi}=\frac{1}{\sqrt{d}}\sum_{i=1}^d\ket{i}\ket{i}$, direct calculations yield
\be
D_q(\ket{\Psi})=\frac{1-d^{1-q}}{q-1}\leqslant \frac{q (d-1)}{d (q-1)},\nonumber
\ee
from which one can show that $D_1(\ket{\Psi})=\ln d$ and 
$D_2(\ket{\Psi})=1-\frac{1}{d}$, as expected. As a second example, we consider a two-qbit state with maximum marginals, $\varrho_{\vec{c}}=\frac{1}{4}(\mathbbm{1}+\sum_{i=1}^3c_i\sigma_i^X\otimes\sigma_i^Y)$, $c_i\in\mathbb{R}$. Using the spectral decomposition $\rho=\sum_i\lambda_i\ket{i}\bra{i}$, where $\lambda_i(c_j)\in[0,1]$, with ordering $\lambda_i\geqslant \lambda_j$ for $i>j$, and a well-known maximization algorithm~\cite{luo08} we obtain
\begin{subequations}
\be
&&D_q(\varrho_{\vec{c}})=\frac{1}{q-1}\left[\sum_{i=1}^4\lambda_i^q-\frac{(1+c)^q+(1-c)^q}{2^{2q-1}}\right],\quad\nonumber\\
&&D_q^{ub}(\varrho_{\vec{c}})=\frac{q}{q-1}\left[\sum_{i=1}^4\left(\lambda_i-\text{\scriptsize $\frac{1}{4}$}\right)\lambda_i^{q-1}-\frac{c_3}{4}\Lambda\right],\nonumber
\ee
\end{subequations}
where $c=\max\{|c_1|,|c_2|,|c_3|\}$ and $\Lambda=\lambda_4^{q-1}+\lambda_3^{q-1}-(\lambda_2^{q-1}+\lambda_1^{q-1})$. The limits $q\to 1,2$ lead to the known results~\cite{celeri11}, and the upper bound has been numerically verified to be never violated for $\varrho_{\vec{c}}$. It is interesting to note that as $q$ varies, the entropic discord $D_1$ continuously deforms into the geometric discord $D_2$ and goes beyond, until reaching $D_{\infty}=0$. This general property of $D_q$ is illustrated in Fig.~\ref{fig1} for two specializations of $\vec{c}$. Concerning the ordering of $D_q$, there is not a typical scenario. For instance, Fig.~\ref{fig1}-(a) shows that $D_q(\varrho_{\vec{c}(v)})>D_q(\varrho_{\vec{c}(v')})$ whenever $D_1(\varrho_{\vec{c}(v)})>D_1(\varrho_{\vec{c}(v')})$, for $v>v'$, whereas in Fig.~\ref{fig1}-(b) an inversion of this ordering is observed around $D_1=0.06$.

%----------------------------------------------------------------------------
\section{On the physics of $q$}

The conceptual framework defined by the Tsallis entropy $S_q$ and its underlying physical implications is supported by a substantial literature. In particular, considerable effort has been endeavored to generalize the laws of thermodynamics (quantum counterparts included)~\cite{plastino01,tsallis95,rajagopal03} to arbitrary values of $q$. In parallel, a significant number of complex systems has been identified whose properties {\em cannot} be suitably described by the celebrated Boltzmann-Gibbs statistical mechanics, in which cases the Tsallis nonextensive statistical mechanics emerges as a very successful generalization~(see Refs.~\cite{oliveira02,tsallis_review} and references therein). Nevertheless, to the best of our knowledge, the precise (uncontroversial) physical meaning of the parameter $q$ remains as a challenging question in the field of the nonextensive statistical mechanics. 

As far as our results are concerned, two positions can be taken. On one hand, we can assume that $S_q$ and $D_q$ are nothing but mathematical deformations of their standard counterparts $S_1$ and $D_1$. According to this view, no further physical content should be added to these generalized quantities, as they would be just different mimics of the same resources. If, on the other hand, the aforementioned context is not neglected, and we believe it should not, then we are invited to ascribe broader significance to the $q$ discord. According to this position, this measure consists in a natural generalization of its embryonic version $D_1$ to the context of complex systems~\cite{oliveira02,tsallis_review}, where the underlying microscopic dynamics dictates the value of $q$. In this sense, the geometric discord $D_2$ turns out to be {\em the} proper measure for the physics defined by $q=2$, rather than just a convenient {\em alternative} to $D_1$. 

Figure \ref{fig1}-(b) illustrates one of the subtleties of an eventual nonextensive theory of discord. For $q=1.5$ unusual behavior is observed, namely, $D_{1.5}$ may become greater than the dimensionless thermodynamic work $D_1=\beta W$. Operationally, this means that in this regime some states may offer optimal informational content. The question then arises about what physical systems can generate such favorable conditions. It should be noted, however, that information gain, $D_q>D_1$, does not immediately imply work gain, $\Delta\mathcal{W}_q>\Delta W$, because the very generalized notions of work $\mathcal{W}_q$ and thermal energy $k_q\mathcal{T}_q$, along with pertinent thermodynamic relations, still wait to be formulated for arbitrary $q$~\cite{plastino01,tsallis95,rajagopal03}. This suggests an interesting research program on the relation between generalized quantum correlations and nonextensive statistical mechanics.

%----------------------------------------------------------------------------
\section{Conclusion}

We have shown that general measures of quantum correlations can be conceived by taking the deviation from Bayes' rule as the only fundamental principle. The resulting measures admit operational interpretation: They indicate by how much quantum demons are more efficient than classical ones in providing information about a given state. Remarkably, our approach reveals a subtle net relation among thermodynamics, quantum correlations, and Bayes' rule. In addition, by choosing a particular entropic principle, we have defined a generalized discord which reproduces the usual entropic and geometric measures by a proper adjustment of a single dimensionless parameter. As a consequence, physical meaning has been given to the geometric discord within the context of complex systems. Besides offering a unified framework for several measures of quantum correlations, our approach opens the venue for challenging explorations of the information-work duet in the field of nonextensive thermodynamics.

%=======================================================================
\section*{Acknowledgments}

This work is supported by CNPq/Brazil and the National Institute for Science and Technology of Quantum Information (INCT-IQ). We acknowledge L. C. Celeri, A. D. Ribeiro, and M. W. Beims for inspiring conversations.

%==========================================================
\appendix
\section{Subsidiary results\label{math}}

Here we prove some of the formulas and results presented throughout the paper. Lemmas 1 and 2 make use of the relations $\Pi_Y[\rho]\,\Pi_y = p_y\rho_{X|y}\Pi_y$ and $\sum_y\Pi_y = \mathbbm{1}_Y$.

%---------------
{\em Lemma 1.} $\text{Tr}[\rho \,g(\Pi_Y[\rho])]=\text{Tr}[\Pi_Y[\rho]\,g(\Pi_Y[\rho])]$.

{\em Proof:} Using the precedent relations and simple properties of projectors ($\Pi_y\Pi_{y'}=\Pi_y\delta_{y,y'}$) we get
\be
\text{Tr}[\rho \,g(\Pi_Y[\rho])] &=& \text{Tr}\sum_y\rho \,g(\Pi_Y[\rho])\,\Pi_y \nonumber \\
&=&  \text{Tr}\sum_y\Pi_y \rho\Pi_y \,g(p_{y}\rho_{X|y})\Pi_{y} \nonumber \\
&=&  \text{Tr}\sum_y\Pi_y \rho\Pi_y \sum_{y'}g(p_{y'}\rho_{X|y'})\Pi_{y'}, \nonumber
\ee
from which the claim follows for any $g$.\hfill{$\blacksquare$}

%---------------
{\em Proposition 1.} {\em $\Delta B(\rho)\geqslant 0$.} 

{\em Proof:} Consider the generalized Klein's inequality~\cite{carlen}, $\text{Tr}[f(A) - f(B)] \geqslant \text{Tr}[(A - B)f'(B)]$, for all differentiable convex function $f:\mathbb{R}\mapsto\mathbb{R}$ and Hermitian matrices $A$ and $B$. Setting $A=\rho$ and $B=\Pi_Y[\rho]$, the positivity in Eq.~\eqref{DB} is proved by use of Lemma 1 with $g=f'$.\hfill{$\blacksquare$}

%---------------
{\em Proposition 2.} {\em $\Delta B(\rho)\leqslant \min_{\Pi_Y}\mathrm{Tr}\big\{\big(\rho-\Pi_Y[\rho]\big)\,f'(\rho)\big\}$.} 

{\em Proof:} Using again the generalized Klein's inequality, take $A=\Pi_Y[\rho]$ and $B=\rho$.\hfill{$\blacksquare$}

%---------------
{\em Proposition 3.} {\em $\Delta B (\rho_X\otimes\rho_Y)=0$.}

{\em Proof:} Taking the eigenstates of $\rho_Y$ as the basis for the measurements we obtain  $\Pi_Y[\rho_Y]=\rho_Y$, from which, by Eq.~\eqref{DB}, it follows that $\Delta B=0$.\hfill{$\blacksquare$}

%---------------
{\em Proposition 4.} {\em $\Delta B(U\rho\,U^{\dag})=\Delta B(\rho)$ for $U=U_X\otimes U_Y$}.

{\em Proof:} Given that $U_{X,Y}$ are unitary operations by hypothesis, then $\text{Tr}f(U\rho\, U^\dag)=\text{Tr}f(\rho)$. In addition,
\be
\text{Tr}f(\Pi_Y[U\rho\, U^\dag])
&=& \text{Tr}f\Big(\sum_yU_Y^{\dag}\Pi_yU_Y\rho\,U^\dag_Y\Pi_yU_Y\Big) \nonumber \\
&=& \text{Tr}f\Big(\sum_y\tilde{\Pi}_y \rho \tilde{\Pi}_y\Big)= \text{Tr} f(\tilde{\Pi}_Y[\rho]),\nonumber
\ee
where $\tilde{\Pi}_y=\ket{\tilde{y}}\bra{\tilde{y}}$ is the projector associated with the new basis $\ket{\tilde{y}}=U_Y\ket{y}$. Taking the minimization over the new basis yields 
\be
\Delta B(U\rho\,U^{\dag})=\min\limits_{\tilde{\Pi}_Y}\text{Tr}\Big(f(\rho)-f(\tilde{\Pi}_Y[\rho])\Big)=\Delta B(\rho).\qquad \blacksquare\nonumber
\ee

%---------------
{\em Definition 1.} {\em We call $\mathcal{F}(\rho)$ continuous if it satisfies the inequality $|\mathcal{F}(\sigma)-\mathcal{F}(\rho)|\leqslant h(\epsilon)$ for all $\rho$, where $h(0)=0$, $\epsilon$ is a small parameter, $\sigma=(1-\epsilon)\rho+\epsilon \tau$ is a perturbation on $\rho$, and $\tau$ is an arbitrary density matrix.}

%---------------
{\em Proposition 5.} {\em $\Delta B(\rho)$ is continuous.} 

{\em Proof:} Using the triangle inequality $|A+B|\leqslant |A|+|B|$ and Eq.~\eqref{DB} we obtain
$|\Delta B(\sigma)-\Delta B(\rho)|\leqslant \Gamma_1+\Gamma_2$, with
\be
&&\Gamma_1\equiv\Big|\text{Tr}\big[f(\sigma)-f(\rho)\big]\Big|,\nonumber \\ &&\Gamma_2\equiv\Big|\max\limits_{\Pi_Y}\text{Tr}\big[f(\Pi_Y[\sigma])-f(\Pi_Y[\rho])\big]\Big|.\nonumber
\ee
Since $\Pi_y[\sigma]=(1-\epsilon)\Pi_Y[\rho]+\epsilon\Pi_Y[\tau]$ and $\text{Tr}f(\rho)$ is continuous by hypothesis, it follows that $\Gamma_{1,2}\leqslant h(\epsilon)$, with $h(0)\!=\!0$. Hence $|\Delta B(\sigma)-\Delta B(\rho)|\leqslant \kappa(\epsilon)$, where $\kappa(\epsilon)\equiv 2h(\epsilon)$ and $\kappa(0)=0$.\hfill{$\blacksquare$}

%---------------
{\em Lemma 2.} $\text{Tr}[(\Pi_Y[\rho])^q]=\sum_yp_y^q\text{Tr}_X[(\rho_{X|y})^q]\,\,(\forall\, q>0)$.

{\em Proof:} Insert $\mathbbm{1}_Y$ in the l.h.s. and take the trace on the subspace $Y$.\hfill{$\blacksquare$}

%---------------
{\em Lemma 3.} $[\text{Tr}_Y(\Pi_y\rho)]^q=\text{Tr}_Y[(\Pi_y\rho\Pi_y)^q]\,\,(\forall\,q>0)$.

{\em Proof:} First, note that $\text{Tr}_Y\Pi_y^q=\sum_{y'}\bra{y'}\,\Pi_y^q\,\ket{y'}=1$ for all $q>0$. Then
\be
[\text{Tr}_Y(\Pi_y\rho)]^q&=&\Big(\bra{y}\rho\ket{y}\Big)^q=\text{Tr}_Y\Pi_y^q\Big(\bra{y}\rho\ket{y}\Big)^q
\nonumber \\ &=&\text{Tr}_Y\Big(\bra{y}\rho\ket{y}\,\Pi_y\Big)^q=\text{Tr}_Y[(\Pi_y\rho\Pi_y)^q].\nonumber
\ee
Analogously, it can be straightforwardly shown that $[\text{Tr}_Y(\Pi_y\rho_Y)]^q=\text{Tr}_Y[(\Pi_y\rho_Y\Pi_y)^q]\,\,(\forall\,q>0)$. \hfill{$\blacksquare$}

%---------------
{\em Lemma 4.} $\sum_yp_y^q=\text{Tr}_Y[(\Pi_Y[\rho_Y])^q]$.

{\em Proof:} Observe that $p_y=\text{Tr}(\Pi_y\rho)=\text{Tr}_Y(\Pi_y\rho_Y)$. Now, use Lemma 3 and the orthogonality of set $\{\Pi_y\rho_Y\Pi_y\}$ to obtain
\be
\sum_yp_y^q=\text{Tr}_Y\sum_y(\Pi_y\rho_Y\Pi_y)^q=\text{Tr}_Y\Big(\sum_y\Pi_y\rho_Y\Pi_y\Big)^q,\nonumber
\ee
from which the claim follows.\hfill{$\blacksquare$}

%---------------
{\em Theorem 1.} {\em ($q$ version of the joint entropy theorem)} $S_q(\Pi_Y[\rho])=S_q(\Pi_Y[\rho_Y])+\sum_yp_y^qS_q(\rho_{X|y})\,\,\,(\forall\,q>0)$.

{\em Proof:} By definition \eqref{Sq} and Lemma 4 we first note that $S_q(\Pi_Y[\rho_Y])=(1-\sum_yp_y^q)/(q-1)$. The result then follows by use of Lemma 2 and straightforward manipulations on $S_q(\Pi_Y[\rho])$.\hfill{$\blacksquare$}

%---------------
{\em Lemma 5.} {\em $\mathrm{Tr}[(\Pi_Y[\rho])^q]\leqslant\mathrm{Tr}_Y\rho_Y^q$ for $\rho$ pure and $q>0$.} 

{\em Proof:} From the definition of conditional state we obtain $\rho_{X|y}=\bra{y}\rho\ket{y}/p_y=\ket{\psi_{X|y}}\bra{\psi_{X|y}}$, where $\ket{\psi_{X|y}}\equiv\langle y|\psi\rangle/\sqrt{p_y}$. Since $\rho_{X|y}$ is pure, $\text{Tr}_X\rho_{X|y}^q=1$, and $\text{Tr}_X\bra{y}\rho\ket{y}^q=p_y^q$. By Lemmas 3 and 4 it follows that
\be
\text{Tr}_Y[(\Pi_Y[\rho_Y])^q]&=&\sum_y p_y^q=\text{Tr}_X\sum_y\bra{y}\rho\ket{y}^q\nonumber \\ &=&\text{Tr}_X\sum_y[\text{Tr}_Y(\Pi_y\rho)]^q=\text{Tr}\sum_y(\Pi_y\rho\Pi_y)^q \nonumber \\
&=& \text{Tr}[(\Pi_Y[\rho])^q].\nonumber
\ee
By Theorem 6 of Ref.~\cite{rastegin} we have $S_q(\Pi_Y[\rho_Y])\geqslant S_q(\rho_Y)$, which implies that $\text{Tr}_Y[(\Pi_Y[\rho_Y])^q]\leqslant \text{Tr}_Y\rho_Y^q$. The above result completes the proof.\hfill{$\blacksquare$}

%==========================================================

\end{document}